\PassOptionsToPackage{unicode}{hyperref}
\PassOptionsToPackage{hyphens}{url}
\PassOptionsToPackage{dvipsnames,svgnames,x11names}{xcolor}
\documentclass[10pt,a4paper,onecolumn]{article}
\usepackage{marginnote}
\usepackage[rgb,svgnames]{xcolor}
\usepackage{authblk,etoolbox}
\usepackage{titlesec}
\usepackage{calc}
\usepackage{tikz}
\usepackage{hyperref}
\hypersetup{%
    unicode=true,
    pdftitle={hcd\_priya: A High-Column-Density Absorber Emulator for
Lyman-α Forest Cosmology},
    pdfauthor={Ming-Feng Ho, Shreyanshi Garg, Camille Avestruz},
    colorlinks=true,
    linkcolor=[rgb]{0.0, 0.5, 1.0},
    citecolor=Blue,
    urlcolor=[rgb]{0.0, 0.5, 1.0},
    breaklinks=true
}
\usepackage{caption}
\usepackage{orcidlink}
\usepackage{tcolorbox}
\usepackage{amssymb,amsmath}
\usepackage{ifxetex,ifluatex}
\usepackage{seqsplit}
\usepackage{xstring}

\makeatletter
\@ifundefined{KOMAClassName}{%
  \IfFileExists{parskip.sty}{%
    \usepackage{parskip}
  }{%
    \setlength{\parindent}{0pt}
    \setlength{\parskip}{6pt plus 2pt minus 1pt}}
}{%
  \KOMAoptions{parskip=half}}
\makeatother
\usepackage{graphicx}
\makeatletter
\newsavebox\pandoc@box
\newcommand*\pandocbounded[1]{%
  \sbox\pandoc@box{#1}%
  \Gscale@div\@tempa{\textheight}{\dimexpr\ht\pandoc@box+\dp\pandoc@box\relax}%
  \Gscale@div\@tempb{\linewidth}{\wd\pandoc@box}%
  \ifdim\@tempb\p@<\@tempa\p@\let\@tempa\@tempb\fi%
  \ifdim\@tempa\p@<\p@\scalebox{\@tempa}{\usebox\pandoc@box}%
  \else\usebox{\pandoc@box}%
  \fi%
}
\def\fps@figure{htbp}
\makeatother
\NewDocumentCommand\citeproctext{}{}
\NewDocumentCommand\citeproc{mm}{%
  \begingroup\def\citeproctext{#2}\cite{#1}\endgroup}
\makeatletter
 \let\@cite@ofmt\@firstofone
 \def\@biblabel#1{}
 \def\@cite#1#2{{#1\if@tempswa , #2\fi}}
\makeatother
\newlength{\cslhangindent}
\newlength{\csllabelwidth}
\newenvironment{CSLReferences}[2] %
 {\begin{list}{}{%
  \setlength{\itemindent}{0pt}
  \setlength{\leftmargin}{0pt}
  \setlength{\parsep}{0pt}
  \ifodd #1
   \setlength{\leftmargin}{\cslhangindent}
   \setlength{\itemindent}{-1\cslhangindent}
  \fi
  \setlength{\itemsep}{#2\baselineskip}}}
 {\end{list}}
\usepackage{calc}

\ifLuaTeX
\usepackage[bidi=basic,shorthands=off,]{babel}
\else
\usepackage[bidi=default,shorthands=off,]{babel}
\fi
\ifLuaTeX
  \usepackage{selnolig} %
\fi
\usepackage{float}
\let\origfigure\figure
\let\endorigfigure\endfigure
\renewenvironment{figure}[1][2] {
    \expandafter\origfigure\expandafter[H]
} {
    \endorigfigure
}

\usepackage[top=3.5cm, bottom=3cm, right=1.5cm, left=1.0cm,
            headheight=2.2cm, reversemp, includemp, marginparwidth=4.5cm]{geometry}

\titleformat{\section}
  {\normalfont\sffamily\Large\bfseries}
  {}{0pt}{}
\titleformat{\subsection}
  {\normalfont\sffamily\large\bfseries}
  {}{0pt}{}
\titleformat{\subsubsection}
  {\normalfont\sffamily\bfseries}
  {}{0pt}{}
\titleformat*{\paragraph}
  {\sffamily\normalsize}

\usepackage{fancyhdr}
\makeatletter
\let\ps@plain\ps@fancy
\fancyheadoffset[L]{4.5cm}
\fancyfootoffset[L]{4.5cm}

\definecolor{linky}{rgb}{0.0, 0.5, 1.0}

\newtcolorbox{repobox}
   {colback=red, colframe=red!75!black,
     boxrule=0.5pt, arc=2pt, left=6pt, right=6pt, top=3pt, bottom=3pt}

\newcommand{\ExternalLink}{%
   \tikz[x=1.2ex, y=1.2ex, baseline=-0.05ex]{%
       \begin{scope}[x=1ex, y=1ex]
           \clip (-0.1,-0.1)
               --++ (-0, 1.2)
               --++ (0.6, 0)
               --++ (0, -0.6)
               --++ (0.6, 0)
               --++ (0, -1);
           \path[draw,
               line width = 0.5,
               rounded corners=0.5]
               (0,0) rectangle (1,1);
       \end{scope}
       \path[draw, line width = 0.5] (0.5, 0.5)
           -- (1, 1);
       \path[draw, line width = 0.5] (0.6, 1)
           -- (1, 1) -- (1, 0.6);
       }
   }

\definecolor{c53baa1}{RGB}{83,186,161}
\definecolor{c202826}{RGB}{32,40,38}

\patchcmd{\@maketitle}{center}{flushleft}{}{}
\patchcmd{\@maketitle}{center}{flushleft}{}{}
\patchcmd{\@maketitle}{\LARGE}{\LARGE\sffamily}{}{}
\def\maketitle{{%
  
  \AB@maketitle}}
\renewcommand\AB@affilsepx{ \protect\Affilfont}
\renewcommand\AB@affilnote[1]{{\bfseries #1}\hspace{3pt}}
\renewcommand{\affil}[2][]%
   {\newaffiltrue\let\AB@blk@and\AB@pand
      \if\relax#1\relax\def\AB@note{\AB@thenote}\else\def\AB@note{#1}%
        \setcounter{Maxaffil}{0}\fi
        \begingroup
        \let\href=\href@Orig
        \let\protect\@unexpandable@protect
        \def\thanks{\protect\thanks}\def\footnote{\protect\footnote}%
        \@temptokena=\expandafter{\AB@authors}%
        {\def\\{\protect\\\protect\Affilfont}\xdef\AB@temp{#2}}%
         \xdef\AB@authors{\the\@temptokena\AB@las\AB@au@str
         \protect\\[\affilsep]\protect\Affilfont\AB@temp}%
         \gdef\AB@las{}\gdef\AB@au@str{}%
        {\def\\{, \ignorespaces}\xdef\AB@temp{#2}}%
        \@temptokena=\expandafter{\AB@affillist}%
        \xdef\AB@affillist{\the\@temptokena \AB@affilsep
          \AB@affilnote{\AB@note}\protect\Affilfont\AB@temp}%
      \endgroup
       \let\AB@affilsep\AB@affilsepx
}
\makeatother

\renewcommand\Affilfont{\sffamily\small\mdseries}
\usepackage{fontspec}
\defaultfontfeatures{Scale=MatchLowercase}
\defaultfontfeatures[\sffamily]{Ligatures=TeX}
\defaultfontfeatures[\rmfamily]{Ligatures=TeX,Scale=1}

\ifPDFTeX\else
\fi
\IfFileExists{upquote.sty}{\usepackage{upquote}}{}
\IfFileExists{microtype.sty}{%
  \usepackage[]{microtype}
  \UseMicrotypeSet[protrusion]{basicmath} %
}{}

\usepackage{fontsetup} %

\PassOptionsToPackage{usenames,dvipsnames}{color} %
\ifLuaTeX
  \usepackage{selnolig}  %
\fi

\title{hcd\_priya: A High-Column-Density Absorber Emulator for Lyman-α
Forest Cosmology}

\author[1,2%
\ensuremath\mathparagraph]{Ming-Feng Ho%
  \,\orcidlink{0000-0002-4457-890X}\,%
}
\author[1%
]{Shreyanshi Garg%
}
\author[1,2%
]{Camille Avestruz%
  \,\orcidlink{0000-0001-8868-0810}\,%
}

\affil[1]{Department of Physics, University of Michigan, USA%
}
\affil[2]{Leinweber Institute for Theoretical Physics, University of
Michigan, USA%
}
\affil[$\mathparagraph$]{Corresponding author}
\date{\vspace{-2.5ex}}

\begin{document}
\maketitle

\marginpar{

  \begin{flushleft}
  \sffamily\small

    {\bfseries Software}
  \begin{itemize}
    \setlength\itemsep{0em}
    \item \href{https://github.com/jibanCat/hcd_priya}{\color{linky}{Repository}} \ExternalLink
  \end{itemize}

  \vspace{2mm}
  
    \par\noindent\hrulefill\par

  \vspace{2mm}

  {\bfseries License}\\
  Authors of papers retain copyright and release the work under a Creative Commons Attribution 4.0 International License (\href{https://creativecommons.org/licenses/by/4.0/}{\color{linky}{CC BY 4.0}}).

  \end{flushleft}
}

\section{Summary}\label{summary}

The one-dimensional flux power spectrum (P1D) of the Lyman-\(\alpha\)
forest probes small-scale structure along quasar lines of sight, with
sensitivity to the scale dependence of primordial fluctuations,
alternatives to cold dark matter, and neutrino masses. Nevertheless, the
effects of high-column-density absorbers (HCDs), particularly
Lyman-limit systems (LLSs), can be strongly degenerate with cosmological
signal on the P1D of high-resolution spectra (\citeproc{ref-Ho2026}{Ho
et al., 2026}), and neglecting the effect of LLSs could shift the
cosmology constraint (\citeproc{ref-ChavesMontero2026}{Chaves-Montero et
al., 2026}). The diffuse intergalactic medium (IGM) and HCDs arise from
the same neutral-hydrogen distribution, but are often treated
separately. We introduce hcd\_priya, a neural-network emulator built on
PRIYA galaxy-formation simulations (\citeproc{ref-Bird2023PRIYA}{Bird et
al., 2023}), that jointly predicts HCD-class-dependent P1D and HCD
population statistics.

\begin{figure}
\centering
\includegraphics[width=1\linewidth,height=\textheight,keepaspectratio]{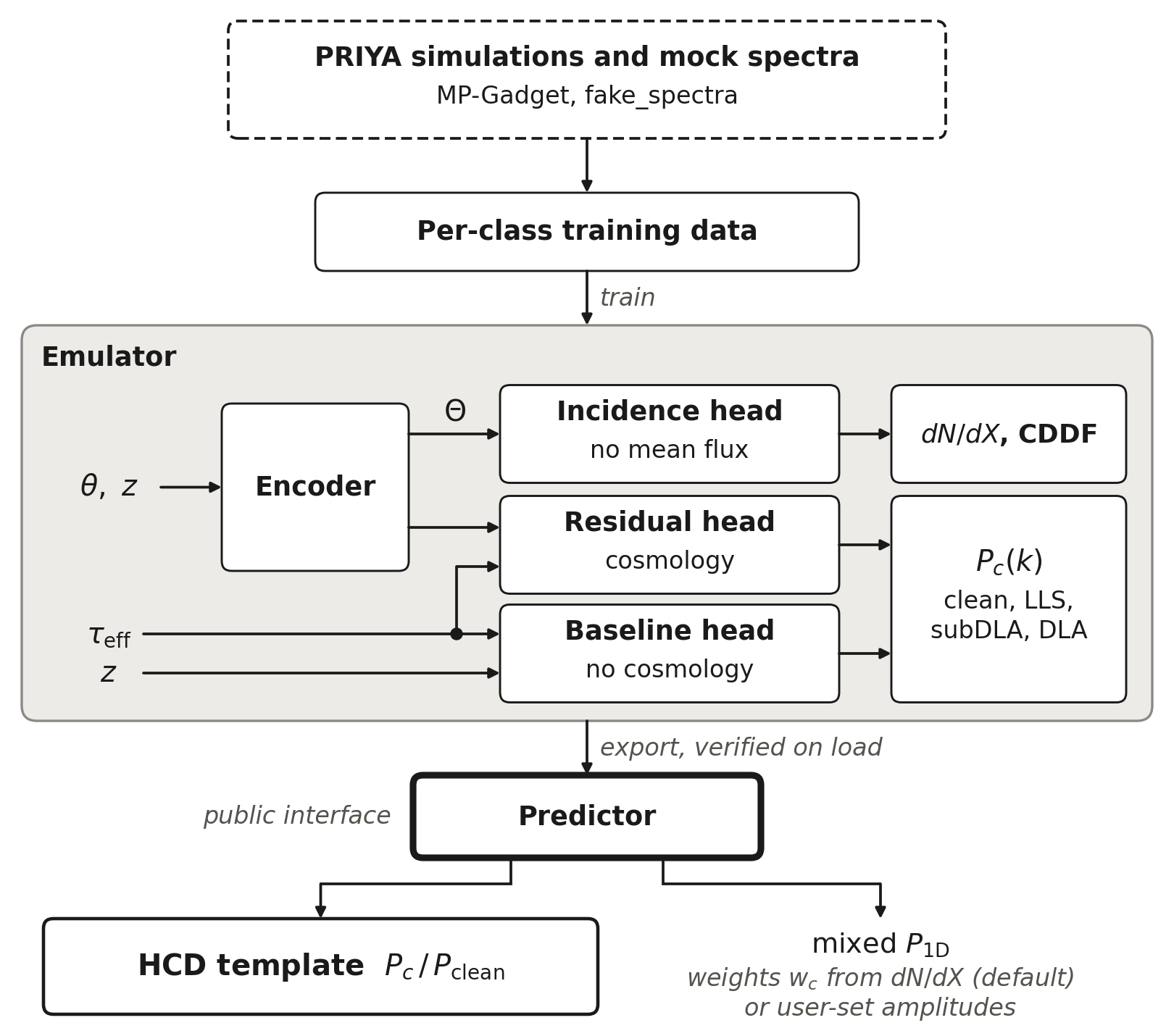}
\caption{\textbf{Software architecture of hcd\_priya.} PRIYA mock
spectra, classified by absorber class, are used to train a shared
encoder with three heads. The predictor combines an ensemble of models
with multi-fidelity corrections. External software: MP-Gadget
(\citeproc{ref-Feng2018MPGadget}{Feng et al., 2018}) and fake\_spectra
(\citeproc{ref-Bird2015}{Bird et al., 2015};
\citeproc{ref-Bird2017FSFE}{Bird, 2017}).\label{fig:schematic}}
\end{figure}

We identify absorbers in simulated sightlines, classify the spectra into
clean, LLS, sub-damped Lyman-\(\alpha\) absorber (subDLA), and damped
Lyman-\(\alpha\) absorber (DLA), and measure the P1D of each HCD class.
Given the cosmological and astrophysical parameters, the hcd\_priya
emulator outputs the P1D of each absorber class, its ratio to the clean
forest, and the absorber incidence and column-density distribution
function (CDDF) (\autoref{fig:schematic}). The emulator can be used to
construct HCD templates for other Lyman-\(\alpha\) emulators or
incorporated into cosmological likelihood analyses using Cobaya
(\citeproc{ref-Torrado2021}{Torrado \& Lewis, 2021}); the JAX forward
model is differentiable, for gradient-based samplers such as NumPyro
(\citeproc{ref-Phan2019NumPyro}{Phan et al., 2019}). The code and the
trained emulator can be installed via \texttt{pip\ install\ hcd-priya}.

\section{Statement of need}\label{statement-of-need}

\begin{figure}
\centering
\includegraphics[width=1\linewidth,height=\textheight,keepaspectratio]{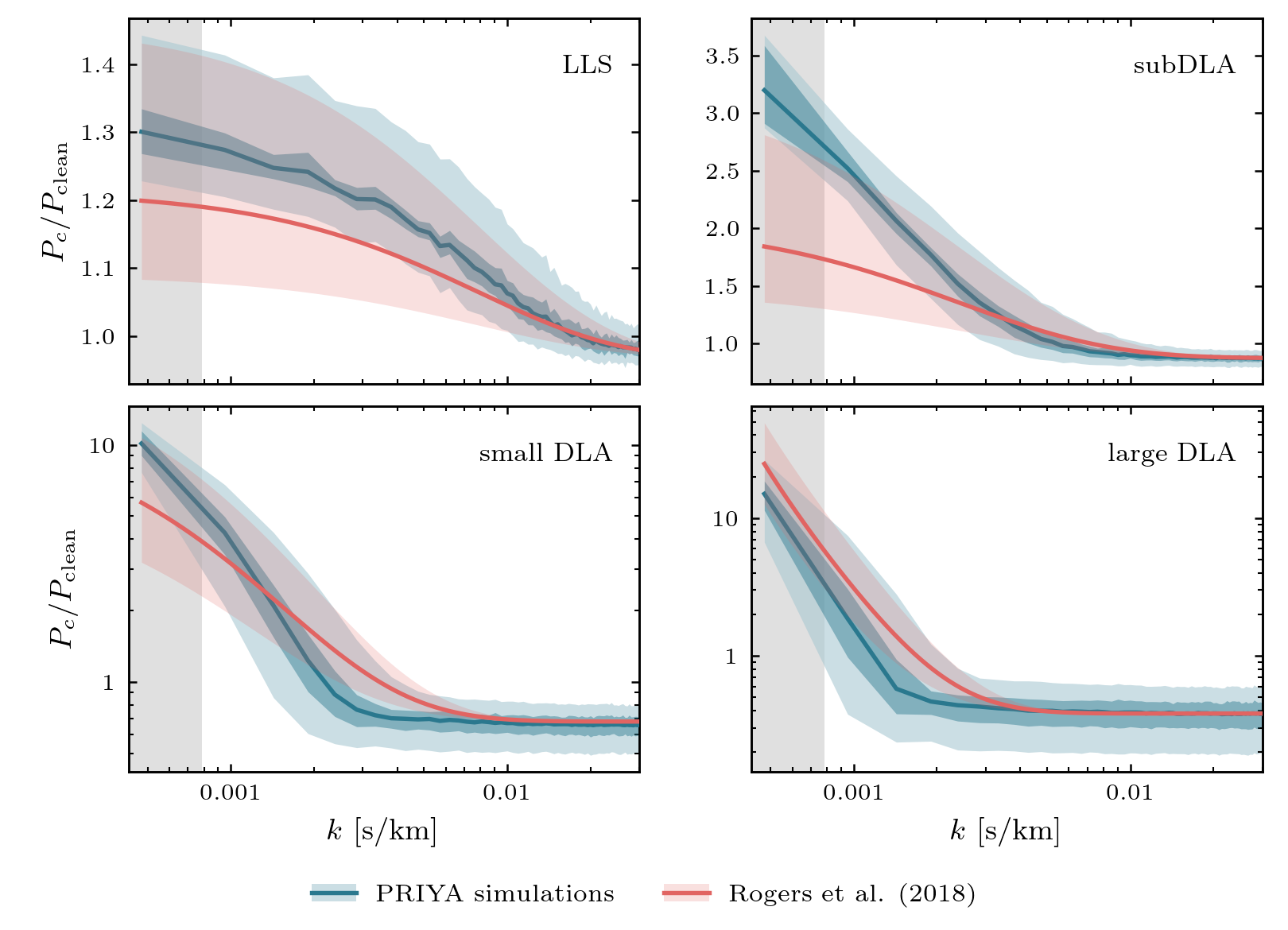}
\caption{\textbf{HCD contributions to the P1D in PRIYA and Rogers et al.
(\citeproc{ref-Rogers2018}{2018}).} The ratio \(P_c/P_{\rm{}clean}\) at
\(z=3.0\), measured with the unfiltered spectra, for the absorber
classes defined in Rogers et al. (\citeproc{ref-Rogers2018}{2018}).
Teal: median of 60 PRIYA simulations; the dark band spans the
simulations, while the light band spans all 20 mean-flux values. Red:
predictions from Rogers et al. (\citeproc{ref-Rogers2018}{2018}), with
amplitudes ranging from 0.5 to 2. Grey: scales beyond the largest mode
of the Rogers simulation.\label{fig:ratio}}
\end{figure}

Cosmological likelihood analyses of the Lyman-\(\alpha\) forest usually
account for the contamination from HCDs using the templates of Rogers et
al. (\citeproc{ref-Rogers2018}{2018}), which are calibrated on a single
Illustris simulation at one cosmology. The Rogers et al.
(\citeproc{ref-Rogers2018}{2018}) templates are useful because in
practice the simulations designed for the diffuse IGM often do not fully
simulate galaxies and their surrounding dense gas associated with HCDs.
Simulating a sufficiently large IGM volume while achieving the
resolution required for the small-scale forest and galaxy formation
remains computationally challenging. However, a template calibrated at
one cosmology does not directly predict how the HCD contribution changes
with cosmological and astrophysical parameters (\autoref{fig:ratio}),
which is often required in a cosmological likelihood analysis.

PRIYA addresses this challenge by combining variations in cosmology and
astrophysics with a full galaxy-formation model
(\citeproc{ref-Bird2023PRIYA}{Bird et al., 2023}). PRIYA's
high-resolution simulations reach a mass resolution comparable to TNG100
(\citeproc{ref-Nelson2019}{Nelson et al., 2019}), which is unusual for a
simulation suite designed for Lyman-\(\alpha\) forest cosmology. By
better resolving the underlying neutral-hydrogen distribution beyond the
IGM, the simulations naturally capture the relative abundances of
different absorber classes and their correlations with the surrounding
forest. hcd\_priya measures the class-dependent P1D and absorber
incidence across the simulation suite and emulates their dependence on
cosmological and astrophysical parameters. We use the predicted
\(dN/dX\) incidence to define the default weights of each absorber
class, bridging the definition of HCD contamination to the physical
definition of the absorber population, rather than treating them
entirely as nuisance parameters.

hcd\_priya is designed for P1D analyses, including DESI DR1
(\citeproc{ref-Karacayli2025}{Karaçaylı et al., 2025};
\citeproc{ref-Ravoux2025}{Ravoux et al., 2025}) and high-resolution
spectra (\citeproc{ref-Ho2026}{Ho et al., 2026}). The emulator also
provides an absorber-cosmology connection baseline for studies of LLS,
subDLA, and DLA.

\section{State of the field}\label{state-of-the-field}

Several tools on the market already model the Lyman-\(\alpha\) forest
P1D: the PRIYA emulators (\citeproc{ref-Bird2023PRIYA}{Bird et al.,
2023}; \citeproc{ref-Fernandez2024}{Fernandez et al., 2024};
\citeproc{ref-Ho2026}{Ho et al., 2026}), LaCE
(\citeproc{ref-CabayolGarcia2023}{Cabayol-Garcia et al., 2023};
\citeproc{ref-Pedersen2021}{Pedersen et al., 2021}), and the Nyx-based
Lyssa emulator (\citeproc{ref-Walther2025}{Walther et al., 2025}). cup1d
provides a cosmological likelihood framework
(\citeproc{ref-ChavesMontero2026}{Chaves-Montero et al., 2026});
effective field theory offers a complementary analytic approach on
quasi-linear scales (\citeproc{ref-Ivanov2024}{Ivanov, 2024}).
Nevertheless, existing P1D emulators do not jointly predict the HCD
population and its cosmology-dependent contribution to the forest; HCDs
are often treated as either contamination or a data-level correction.

We developed hcd\_priya as a standalone package. The advantage is that
hcd\_priya can directly serve as an additional HCD template on top of
existing P1D likelihoods. Also, extending the Ho et al.
(\citeproc{ref-Ho2026}{2026}) Gaussian-process (GP) emulator to multiple
HCD classes and population statistics would be computationally
expensive. We instead train a shared multi-head neural network in JAX
and Equinox (\citeproc{ref-Bradbury2018JAX}{Bradbury et al., 2018};
\citeproc{ref-Kidger2021Equinox}{Kidger \& Garcia, 2021}), as suggested
by the recent developments in neural network multi-fidelity emulation
(\citeproc{ref-Yang2026}{Yang et al., 2026}).

\section{Software design}\label{software-design}

We classify each mock sightline by its highest-column-density absorber,
keeping the full flux field and correlations between systems. With
\(m=\max_i\log_{10}(N_{{\rm{}HI},i}/{\rm{}cm^{-2}})\), where
\(N_{{\rm{}HI},i}\) is the column density of neutral hydrogen absorber
\(i\), the class \(c\) is

\[c=\begin{cases}{\rm{}DLA},&m\ge20.3,\\{\rm{}subDLA},&19.0\le{}m<20.3,\\{\rm{}LLS},&17.2\le{}m<19.0,\\{\rm{}clean},&{\rm{}otherwise}.\end{cases}\]

For each mean flux \(\bar{F}\), we define a common optical-depth scaling
factor \(s\) from the PRIYA DLA-filtered spectra
(\citeproc{ref-Bird2023PRIYA}{Bird et al., 2023};
\citeproc{ref-Fernandez2024}{Fernandez et al., 2024}), satisfying

\[\left\langle{}e^{-s\tau_{\rm{}filtered}}\right\rangle=\bar{F},\]

where \(\tau_{\rm{}filtered}\) is the optical depth after the DLA filter
and \(\langle\cdot\rangle\) averages over all pixels.

We apply the same mean-flux scaling factor to the filtered spectra of
all absorber classes, and measure their P1D relative to the target mean
flux using fake\_spectra (\citeproc{ref-Bird2015}{Bird et al., 2015};
\citeproc{ref-Bird2017FSFE}{Bird, 2017}):

\[\delta_F(v)=\frac{e^{-s\tau_{\rm{}filtered}(v)}}{\bar{F}}-1,\]

where \(\delta_F(v)\) is the flux contrast at velocity \(v\). The class
P1D, \(P_c(k)\), is the two-point statistic of \(\delta_F\) at
wavenumber \(k\), as defined in fake\_spectra.

The filter cuts on sightlines whose peak optical depth goes beyond
\(10^{6}\), and replaces the absorbed pixels around that peak with the
mean flux.

The low-fidelity emulator maps the nine parameters and redshift to a
shared latent representation \(\Theta\). Three heads predict the
absorber statistics, the P1D baseline, and its parameter-dependent
residual. For each absorber class \(c\), we use a structured mean
(\citeproc{ref-KennedyOHagan2001}{Kennedy \& O'Hagan, 2001}):

\[\ln{}P_c(k)=b_c(k;z,\tau_{\rm{}eff})+\sigma_c(k)\,r_c(k;\Theta,\tau_{\rm{}eff}),\]

where \(z\) is the redshift, \(\tau_{\rm{}eff}=-\ln\bar{F}\) the
effective optical depth, \(b_c\) the baseline, \(r_c\) the residual, and
\(\sigma_c(k)\) the scatter of \(\ln{}P_c\) across the training
simulations. The P1D heads predict coefficients in separate, learned
low-rank bases over wavenumbers, initialized using singular value
decomposition of the training spectra. We average predictions from five
independently trained networks as an ensemble. Using five
high-resolution runs, we apply a multi-fidelity correction to correct
the resolution convergence, similar to Ho et al.
(\citeproc{ref-Ho2026}{2026}):

\[P_c^{\rm{}MF}(k)=P_c^{\rm{}LF}(k)\,e^{g_c},\]

where LF and MF mark the low-fidelity and multi-fidelity outputs, and
\(g_c\) is a multi-fidelity correction function defined in the code.

The predicted absorber incidence, \(dN/dX\), determines the default
weights of each sightline class. For class \(c\), the expected number of
absorbers is

\[\mu_c=\left(\frac{dN}{dX}\right)_c\Delta{}X,\]

where \((dN/dX)_c\) is the number of absorbers (class \(c\)) per unit
distance \(X\), and \(\Delta{}X\) the absorption distance of a
sightline.

Assuming Poisson counts, the classification rule gives

\[\begin{aligned}w_{\rm{}DLA}&=1-e^{-\mu_{\rm{}DLA}},\\w_{\rm{}subDLA}&=\left(1-e^{-\mu_{\rm{}subDLA}}\right)e^{-\mu_{\rm{}DLA}},\\w_{\rm{}LLS}&=\left(1-e^{-\mu_{\rm{}LLS}}\right)e^{-\mu_{\rm{}subDLA}-\mu_{\rm{}DLA}},\\w_{\rm{}clean}&=e^{-\sum_c\mu_c},\end{aligned}\]

where \(w_c\) is the fraction of sightlines in class \(c\); the sum in
\(w_{\rm{}clean}\) is over the three HCD classes.

A separate fitted redshift correction accounts for departures from the
Poisson approximation. The weights reconstruct the total P1D through
\(P_{\rm{}total}=\sum_cw_cP_c\), and users can vary the HCD amplitudes
in cosmological likelihood analyses. The predictor flags extrapolation
beyond the training range.

\section{Research impact statement}\label{research-impact-statement}

hcd\_priya is a continuation of our work in Ho et al.
(\citeproc{ref-Ho2026}{2026}), where we found a strong degeneracy
between LLS abundance and cosmological parameters inferred from the
small-scale Lyman-\(\alpha\) forest. The HCD catalogues output from
hcd\_priya are currently being analyzed by S. Garg et al.~(in prep.) to
investigate how different physical prescriptions and cosmological models
influence the distribution of neutral hydrogen, and to assess the impact
of HCDs as contaminants in the Lyman-\(\alpha\) forest. We also plan to
revisit high-resolution P1D constraints using this simulation-informed
HCD prior constructed from hcd\_priya. As a public package, hcd\_priya
provides an HCD model that can be incorporated into existing
cosmological analyses.

\begin{figure}
\centering
\includegraphics[width=1\linewidth,height=\textheight,keepaspectratio]{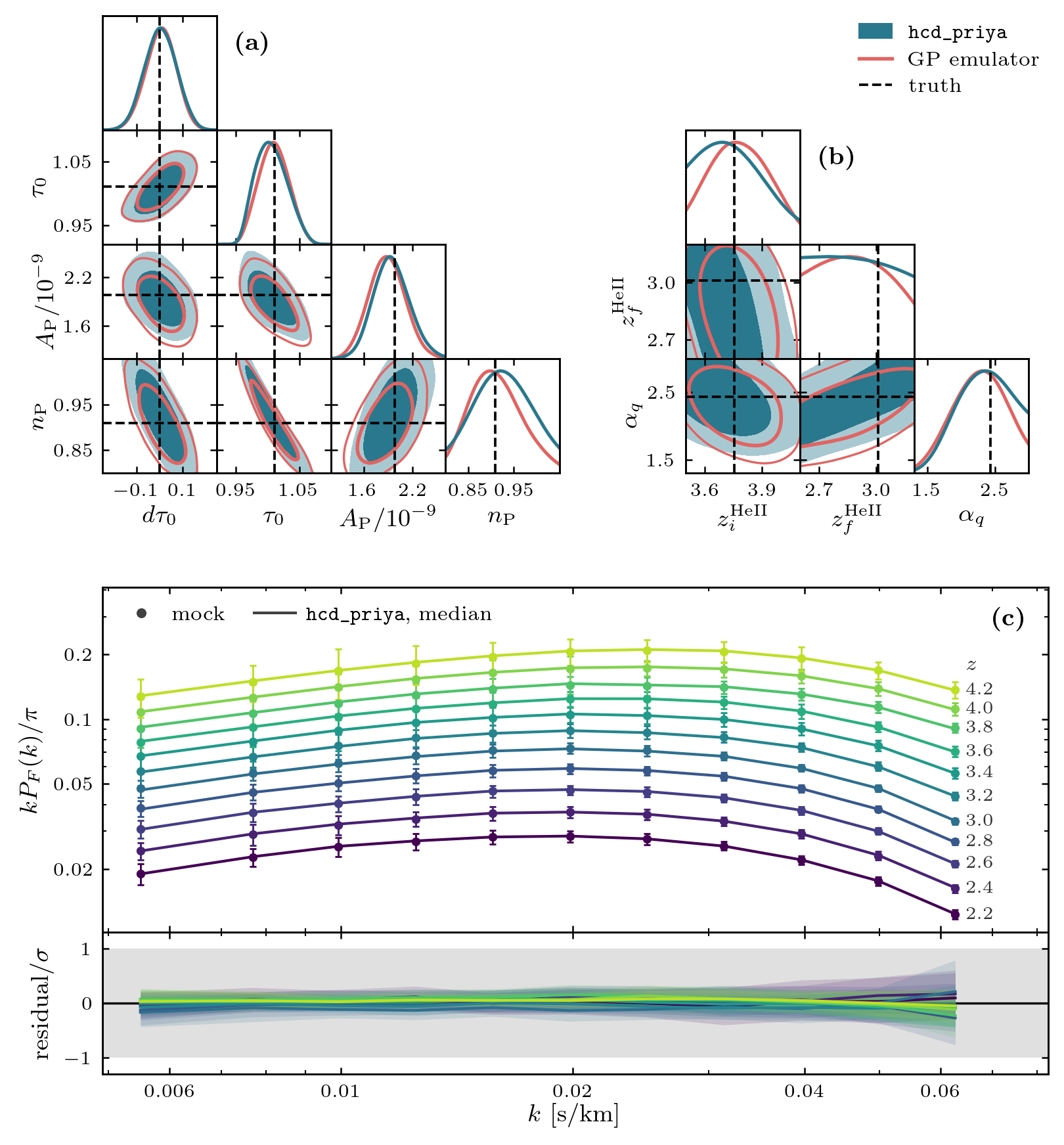}
\caption{\textbf{Cosmological parameter recovery from a held-out PRIYA
simulation.} (a, b) Posterior constraints from hcd\_priya (filled 68\%
and 95\% regions) and the PRIYA GP emulator (red contours; the GP was
trained on this simulation). Dashed lines indicate the input parameters.
(c) The mock P1D and its posterior median prediction, with residuals in
units of the observational uncertainty. The GP chain is not publicly
available.\label{fig:mock}}
\end{figure}

The mock-recovery example in the repository uses the noiseless P1D from
a held-out PRIYA high-resolution simulation, computed at 121
KODIAQ-SQUAD bins and covariance (\citeproc{ref-Karacayli2022}{Karaçaylı
et al., 2022}; \citeproc{ref-Murphy2019}{Murphy et al., 2019};
\citeproc{ref-OMeara2015}{O'Meara et al., 2015},
\citeproc{ref-OMeara2017}{2017}). \autoref{fig:mock} compares the
hcd\_priya reference posterior with that of the GP emulator of Ho et al.
(\citeproc{ref-Ho2026}{2026}) on the same mock. Both chains are
statistically consistent: the posterior means differ by at most 0.27
posterior standard deviations, with posterior width ratios between 1.00
and 1.09. Both fits use KODIAQ-SQUAD data covariance alone and keep the
HCD amplitudes fixed. We chose to leave HCD marginalization out of the
likelihood because the cosmological analysis using hcd\_priya belongs to
a separate analysis paper. The mock, the model trained without that
simulation, and the stored reference are distributed with the software
release.

\begin{figure}
\centering
\includegraphics[width=1\linewidth,height=\textheight,keepaspectratio]{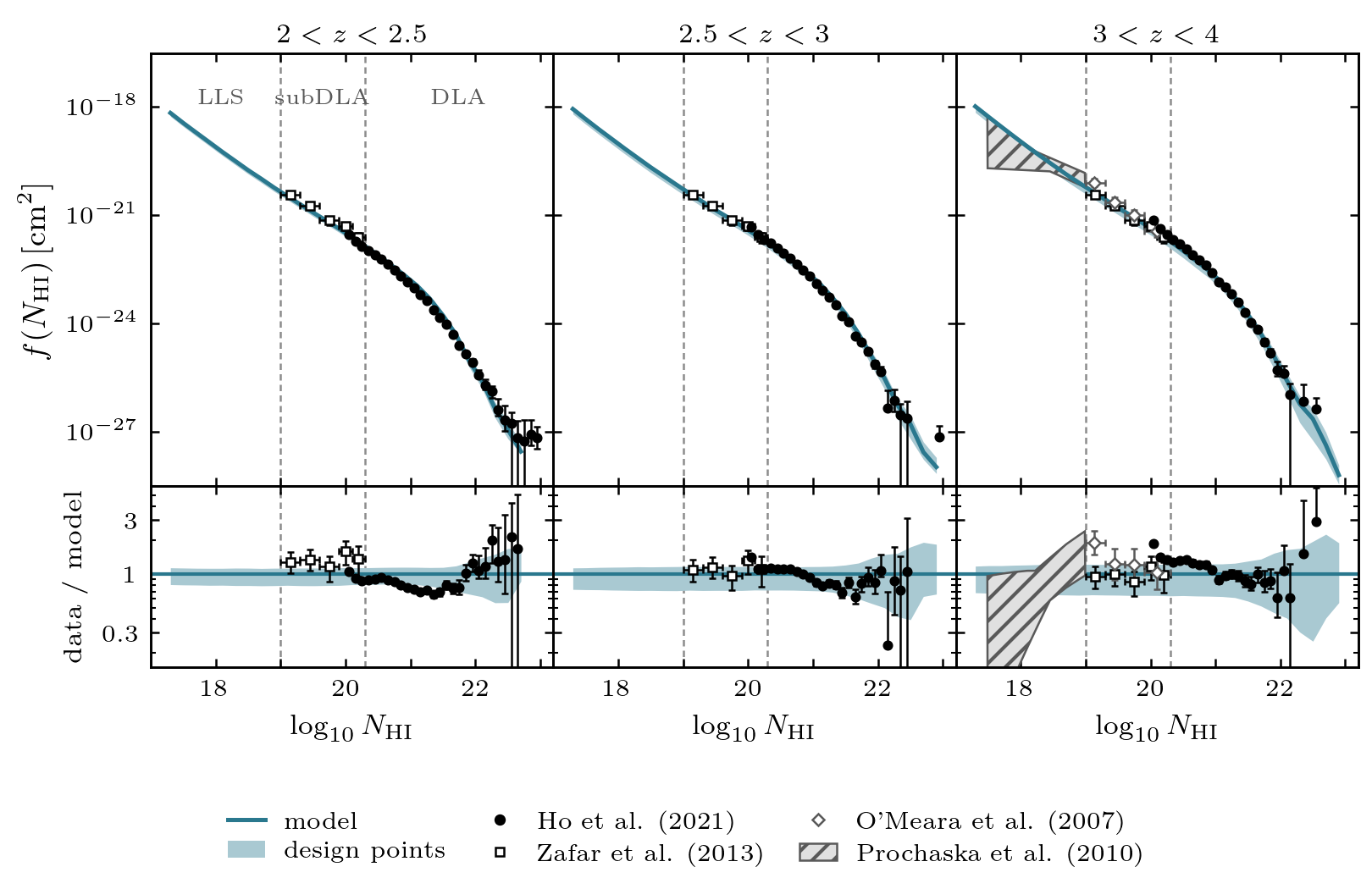}
\caption{\textbf{Predicted HCD column-density distribution compared with
observations.} The emulated CDDF, \(f(N_{\rm{}HI})\), at three redshifts
is compared with observations: DLAs from Ho et al.
(\citeproc{ref-HoBirdGarnett2021}{2021}), subDLAs from Zafar et al.
(\citeproc{ref-Zafar2013}{2013}) and O'Meara et al.
(\citeproc{ref-OMeara2007}{2007}), and LLSs from Prochaska et al.
(\citeproc{ref-Prochaska2010}{2010}). Bands span the 60 design points of
parameter variations. Dashed lines indicate the absorber class
boundaries; the lower panels show the ratio of observations to the
model.\label{fig:cddf}}
\end{figure}

On the absorber population side, we compare the predicted CDDF with
published measurements of LLSs, subDLAs, and DLAs (\autoref{fig:cddf}).
The CDDFs are evaluated at the midpoint of the PRIYA parameter prior
ranges, without fitting to observational data points.

\section{AI usage disclosure}\label{ai-usage-disclosure}

The authors thank Claude (Fable 5 and 5.2, Opus 4.7, 4.8, and 5.5, and
Sonnet 4.6) through Claude Code, GPT (GPT 5.6 Sol and GPT 6 Astra), and
GitHub Copilot. These tools contributed substantially to the
implementation, testing, documentation, and code review of hcd\_priya.
They also assisted with the manuscript: expanding the human-written
draft; checking references and numerical results; and developing the
plotting code. The authors defined the scientific questions, emulator
architecture, validation metrics, and inference workflow. The authors
evaluated alternative approaches and made the final scientific
decisions. The authors reviewed, edited, and validated all AI-assisted
outputs using automated tests and comparisons with analytical
expectations, existing emulators, and existing scientific software. The
authors take responsibility for the software and manuscript.

\section{Acknowledgements}\label{acknowledgements}

The authors thank Simeon Bird and the PRIYA simulation group for sharing
the simulated sightlines of PRIYA. MFH thanks Alexandra Wells for useful
discussions on DLA clustering, and thanks Yanhui Yang and Mahdi (Sum)
Qezlou for useful discussions on neural nets. MFH and CA are supported
by the Leinweber Foundation and DOE grant DE-SC0019193. This research
was supported in part through computational resources and services
provided by Advanced Research Computing (ARC), a division of Information
and Technology Services (ITS) at the University of Michigan, Ann Arbor.
The PRIYA simulations were run on Frontera at the Texas Advanced
Computing Center (LRAC AST21005), made possible by National Science
Foundation award OAC-1818253. This work made use of NumPy
(\citeproc{ref-Harris2020NumPy}{Harris et al., 2020}), SciPy
(\citeproc{ref-Virtanen2020SciPy}{Virtanen et al., 2020}), h5py
(\citeproc{ref-Collette2013h5py}{Collette, 2013}), pandas
(\citeproc{ref-McKinney2010pandas}{McKinney, 2010}), Matplotlib
(\citeproc{ref-Hunter2007Matplotlib}{Hunter, 2007}), and Optax
(\citeproc{ref-DeepMind2020JAX}{DeepMind et al., 2020}).

\section*{References}\label{references}
\addcontentsline{toc}{section}{References}

\protect\phantomsection\label{refs}

\end{document}